\documentclass{article}
\usepackage{ijcai21}

\usepackage{times}
\usepackage{soul}
\usepackage{url}
\usepackage[hidelinks]{hyperref}
\usepackage[utf8]{inputenc}
\usepackage[small]{caption}
\usepackage{graphicx}
\usepackage{amsmath}
\usepackage{amsthm}
\usepackage{booktabs}
\usepackage{algorithm}
\usepackage{algorithmic}
\usepackage{xcolor}
\usepackage{multirow}
\usepackage{amssymb}
\title{Reinforcement Learning Based Sparse Black-box Adversarial Attack on Video Recognition Models}

\author{
Zeyuan Wang\and
Chaofeng Sha\footnote{Corresponding Authors}\And
Su Yang\footnotemark[1]\\
\affiliations
Shanghai Key Laboratory of Intelligent Information Processing, \\School of Computer Science, Fudan University\\
\emails
\{18210440020, cfsha, suyang\}@fudan.edu.cn
}

\begin{document}

\maketitle

\begin{abstract}
  We explore the black-box adversarial attack on video recognition models.
  Attacks are only performed on selected key regions and key frames to reduce the high computation cost of searching adversarial perturbations on a video due to its high dimensionality. 
  To select key frames, one way is to use heuristic algorithms to evaluate the importance of each frame and choose the essential ones. 
  However, it is time inefficient on sorting and searching. 
  In order to speed up the attack process, we propose a reinforcement learning based frame selection strategy. 
  Specifically, the agent explores the difference between the original class and the target class of videos to make selection decisions. 
  It receives rewards from threat models which indicate the quality of the decisions.
  Besides, we also use saliency detection to select key regions and only estimate the sign of gradient instead of the gradient itself in zeroth order optimization to further boost the attack process. 
  We can use the trained model directly in the untargeted attack or with little fine-tune in the targeted attack, which saves computation time. 
  A range of empirical results on real datasets demonstrate the effectiveness and efficiency of the proposed method.
  
\end{abstract}

\section{Introduction}
In recent years, more and more machine learning models, especially deep learning models, are deployed in various vital fields with high security requirements, such as face recognition, speech recognition, video surveillance, etc.
However, these models are vulnerable to adversarial examples~\cite{goodfellow2014explaining}, which are formed by applying slight but worst-case perturbations to examples from the data distribution, and might be misclassified by models with high confidence.
This increases people's interest in adversarial attacks.
Some researchers have investigated resisting adversarial examples~\cite{buckman2018thermometer} to improve model security, while others have explored how to attack networks~\cite{cheng2018query,cheng2019sign,wei2019sparse,wei2020heuristic} to assess their robustness.

Different from the white-box attack in which attackers have access to the model's parameters, we focus on the black-box attack where the model is unknown to attackers in this work.
Jia \emph{et al.}~\shortcite{jia2019identifying} summarize that there are two types of black-box attack methods for video recognition models, dense attack and sparse attack.
In dense attack~\cite{li2018adversarial}, all the frames in a video are polluted.
However, the dimensionality of video data is high since videos have both temporal and spatial dimensions.
Based on the insight that different regions or frames in a video contribute differently to the video classification result, and there contains redundant information in a video when feeding to the model,
Wei \emph{et al.}~\shortcite{wei2019sparse} introduce sparse attack, where only some key frame and some key regions in a video are polluted, which leads to the reduction of adversarial perturbations and the increasing efficiency of attacking.

The major challenge in sparse attack on video recognition models is how to select key regions and key frames.
Wei \emph{et al.}~\shortcite{wei2020heuristic} introduce a heuristic algorithm, in which they evaluate the importance of each salient region and frame, select regions and frames according to the descending order of importance scores, and finally perform attacks on the selected regions of selected frames.
However, the key frames selection process in their methods is not time efficient.
In some conditions such as a video example is close to its decision boundary, the performance of this method can even be worse than dense attack methods.

Reinforcement Learning (RL) has strong potential in sequence decision making.
To further improve the efficiency of video attacking, we propose a RL based sparse black-box adversarial attack method on video recognition models.
Specifically, we model the key frames selection process as a Markov Decision Process (MDP).
The agent in RL explores the differences between the original class and the target class of videos, and determines the initialized direction of optimization to make key frames selection decisions.
The agent receives rewards by querying the threat model which indicate the quality of the current selection decision.
We use a Convolutional Neural Network (CNN) to extract features from the noise vectors and sample frames with a multinomial distribution.
Proximal Policy Optimization (PPO) algorithm~\cite{schulman2017proximal} is applied to optimize the policy.
In particular, we divide the policy optimization into a two-stage approach to accelerate the key frames selection process.
The agent can learn a general policy using video examples from any classes first, and then fine-tune its policy using the video examples from the specific class.
Optimization-based attack method is finally performed to do the attack. 
Besides, we also use the fine grained saliency approach to construct saliency maps to select key regions on each frame and only estimate the sign of gradient instead of the gradient itself when performing attacks using zeroth order optimization to further boost the process.
Our major contributions can be summarized as follows:
\begin{itemize}
\item We model the key frames selection process in sparse video attack as an MDP and design a two-stage training approach for policy optimization. 
\item A query-efficient sparse black-box adversarial attack method on video recognition models is proposed based on saliency detection, key frames selection and Sign-OPT method.
\item Extensive experiments on two benchmark datasets and two widely used video recognition models demonstrate the effectiveness and efficiency of the proposed method. It reduces more than 34\% in query numbers and time compared to the current attack methods.
\end{itemize}

\section{Related Work}
\subsection{Adversarial Attack on Image and Video Models}
Past works on adversarial attack are mostly focused on image recognition models.
Many attack algorithms for image attack have been proposed, such as transfer based attack~\cite{papernot2017practical} and query-based attack~\cite{cheng2018query,cheng2019sign}.

Li \emph{et al.}~\shortcite{li2018adversarial} first extend adversarial attack to video recognition models with a generative dense attack method.
Wei \emph{et al.}~\shortcite{wei2019sparse} propose that we don’t need to pollute every frame and use an iterative optimization algorithm based on $l_{2,1}$-norm to sparsely attack the video classification models successfully.
Zajac \emph{et al.}~\shortcite{zajac2019adversarial} introduce spatial sparsity to adversarial attack by only adding an adversarial framing on the border of each frame.
However, these works are all white-box attacks.
Jiang \emph{et al.}~\shortcite{jiang2019black} propose the first black-box video attack framework which utilizes tentative perturbations and partition-based rectifications to obtain good adversarial gradient estimates with fewer queries.
Sparse attack is introduced to black-box video attack in~\cite{wei2020heuristic}, where they evaluate the importance of each salient region and each frame, heuristically search a subset of regions and frames, and only perform attacks on the selected subset.
However, they consume too much queries sorting and searching, which offsets the advantages of sparsity.
The recent arXiv paper~\cite{yan2020sparse} first introduces RL to temporal sparse black-box video attack and they combine the key frames selection with the attacking step as a whole process.
But they train agents for each model and use complex networks to extract features which is still not efficient enough.

Our method is also a sparse attack method. But unlike the algorithms mentioned above, the agent in our method only explores the difference between the original class and the target class of videos to make selection decisions and receives rewards from the threat model directly. It can be trained with a two-stage approach, and is much more query-efficient. 

\subsection{Deep Reinforcement Learning}
Deep RL~\cite{mnih2013playing} is a category of artificial intelligence where agents can get rewards or punishments based on their actions, which is similar to the way humans learn from experience.
It has received a lot of attention since AlphaGo~\cite{silver2016mastering} beats humans, including those who work on adversarial attack problems.
Lin \emph{et al.}~\shortcite{lin2017tactics} introduce tactics to attack deep RL agents, which is distinct from our work.
Yang \emph{et al.}~\shortcite{yang2020patchattack} use deep RL to generate adversarial image examples.
However, except the recent arXiv paper~\cite{yan2020sparse}, there are no other examples that deep RL is applied to attack video recognition models.

In computer vision domain, Tang \emph{et al.}~\shortcite{tang2018deep} use a deep progressive RL method to distil the most informative frames in sequences for recognizing actions.
Zhou \emph{et al.}~\shortcite{zhou2018deep} formulate video summarization as a sequential decision-making process and develop a deep RL network to summarize videos.
However, these works focus more on the inner features of a video and are different from ours.

\section{Methodology}
In this section, we introduce the proposed sparse adversarial attack method.
For a soft-label black-box attack, we can only get the predicted top one class label and its probability from the video recognition model when given a clean video.

Denote the video recognition model as a function $F$.
Specifically, it takes a clean video $\boldsymbol{x}\in \mathbb{R}^{T\times W\times H\times C}$ as an input and outputs the predicted top one class label $y$ and its probability $p(y|\boldsymbol{x})$. 
Here $T$, $W$, $H$ and $C$ denote the length (i.e., the number of frames), width, height and the number of channels of the video, respectively. 
If the prediction is correct, then $y$ equals the ground-truth label $\bar{y}$.
Our goal is to generate an adversarial example $\boldsymbol{x}_\text{adv}$ close to $\boldsymbol{x}$ such that $F(\boldsymbol{x}_\text{adv})\neq\bar{y}$ in the untargeted attack or $F(\boldsymbol{x}_\text{adv})=y_\text{adv}$ in the targeted attack, where $y_\text{adv}$ is the targeted adversarial class label. We introduce a mask $\boldsymbol{M}\in\{0,1\}^{T\times W\times H\times C}$ to represent temporal and spatial sparsity.
For each element in $\boldsymbol{M}$, only if its value is equal to $1$, then the corresponding pixel in the video will be included in the perturbation process.

\subsection{Saliency Detection for Spatial Sparsity}

We apply spatial sparsity to each frame of the video by the saliency maps generation approach. 
This idea is from both the human visual system and some image and video classification models that focused locations contribute more to the recognition results.

We use the fine grained saliency approach ~\cite{montabone2010human} for saliency maps construction.
It calculates saliency based on center-surround differences.
We control the area ratio of the salient regions in one frame by introducing a parameter $\varphi\in(0,1]$.
Larger $\varphi$ leads to smaller spatial sparsity.
Denote the fine grained saliency detection as a function $S$, it initializes $\boldsymbol{M}$ by setting the values of the salient regions for all video frames to 1.
The initialized mask will be
\begin{equation}
     \boldsymbol{M}_\text{spatial}=\left\{
\begin{aligned}
& S(\boldsymbol{x},\varphi) && \text{(U)} \\
& S(\boldsymbol{x},\varphi) \cup S(\hat{\boldsymbol{x}},\varphi) && \text{(T),}
\end{aligned}
\right.
\end{equation}
where (U) and (T) represent “in untargeted attack” and “in targeted attack”, respectively.
$\hat{\boldsymbol{x}}$ is a video sample of target class $y_{adv}$ here in targeted attack and will be a video sample which does not belong to $\bar{y}$ and set as the “target direction” of attacking in untargeted attack.

\subsection{Reinforcement Learning Based Key Frames Selection for Temporal Sparsity}

For temporal sparsity, we use the “unimportant frames deletion” method to select a subset of frames that contribute the most to the success of an adversarial attack from the successive frames.
We consider the key frames selection process as a multi-step MDP and use RL to learn a policy to select frames.
In this subsection, we first formulate the MDP of key frames selection and then introduce the policy network and optimization algorithm.

\subsubsection{Markov Decision Process for Key Frames Selection}
The MDP consists of states, actions, transition function and reward function.

\paragraph{State.} The state $\boldsymbol{s}_t$ at decision step $t$ is set as the difference between the original class and the target class of videos masked by the current temporal and spatial sparsity mask
\begin{equation}
\boldsymbol{s}_t=(\hat{\boldsymbol{x}}-\boldsymbol{x})\odot \boldsymbol{M}_t.
\end{equation}
Here, $\odot$ represents Hadamard product.
At the initial state $\boldsymbol{s}_0$, the mask is the spatial sparsity mask $\boldsymbol{M}_\text{spatial}$.

\paragraph{Action.} An action $a_t$ is a frame chosen from all $T$ frames that the agent thinks it's not a key frame at decision step $t$.

\paragraph{State transition.} Once the agent chooses action ${a_t}$, we set the values of ${a_t}^{\text{th}}$ frame in $\boldsymbol{M}_t$ to $0$.
Denote this process as a function $D$, then we can update the state to
\begin{equation}
\boldsymbol{s}_{t+1}=(\hat{\boldsymbol{x}}-\boldsymbol{x})\odot \boldsymbol{M}_{t+1}=(\hat{\boldsymbol{x}}-\boldsymbol{x})\odot D(\boldsymbol{M}_t, a_t).
\end{equation}
$\boldsymbol{s}_{t+1}$ will be the terminal state either if $F(\boldsymbol{x}+\boldsymbol{s}_{t+1})=\bar{y}$ in untargeted attack or $F(\boldsymbol{x}+\boldsymbol{s}_{t+1})\neq y_\text{adv}$ in targeted attack
(i.e., the ${a_t}^{\text{th}}$ frame is a key frame),
or $a_t \in \{ a_1, a_2, ..., a_{t-1}\}$ 
(i.e., the ${a_t}^{\text{th}}$ frame has already been chosen in the past decision steps).

\paragraph{Reward.} 
The reward reflects the quality of the chosen action.
Our goal is that the agent learns to delete the least important frame and maximizes the long term expected reward.
Based on this intention, we design the reward during pretraining as
\begin{equation}
\resizebox{.91\linewidth}{!}{$
    \displaystyle
     r_t=\left\{
\begin{aligned}
& -1 && \text{if } a_t \in \{ a_1, a_2, ..., a_{t-1}\} \\
& 0 && \text{else if } F(\boldsymbol{x}+\boldsymbol{s}_{t+1}) = \bar{y} \text{ (U) or } F(\boldsymbol{x}+\boldsymbol{s}_{t+1}) \neq y_\text{adv} \text{ (T)}\\
& 1 && \text{else if } F(\boldsymbol{x}+\boldsymbol{s}_{t+1}) \neq \bar{y}\text{ (U) or }F(\boldsymbol{x}+\boldsymbol{s}_{t+1}) = y_\text{adv}\text{ (T)}.
\end{aligned}
\right.
$}
\end{equation}
We set different rewards for the two terminal states to distinguish between deleting unimportant frames and duplication.
The former one might be a helpless action when no other unimportant frames can be found, while the latter one should not be done at any time.
Note that in targeted attack, for certain $\boldsymbol{x}$ and $\hat{\boldsymbol{x}}$, the larger value of $p(y_\text{adv}|\boldsymbol{x}+\boldsymbol{s}_{t+1})$ means the less importance of ${a_t}^{\text{th}}$ frame toward generating the adversarial example.
Thus we define the reward during fine-tuning as
\begin{equation}
     r_t=\left\{
\begin{aligned}
& -1 && \text{if } a_t \in \{ a_1, a_2, ..., a_{t-1}\} \\
& 0 && \text{else if } F(\boldsymbol{x}+\boldsymbol{s}_{t+1}) \neq y_\text{adv} \\
& p(y_\text{adv}|\boldsymbol{x}+\boldsymbol{s}_{t+1}) && \text{else if } F(\boldsymbol{x}+\boldsymbol{s}_{t+1}) = y_\text{adv}.
\end{aligned}
\right.
\end{equation}

\subsubsection{Policy Network}
The agent behavior is defined by its policy.
We use a policy network $\pi_{\boldsymbol{\theta}_\text{p}}(a_t|\boldsymbol{s}_t)$ to parameterize the policy.
\begin{algorithm}[t]
\caption{Policy optimization for selecting key frames}
\label{alg:algorithm1}
\textbf{Input}: Clean video set $\boldsymbol{X}$, target class video set $\hat{\boldsymbol{X}}$, video recognition model $F$. \\
\textbf{Parameter}: Iteration number $I$, actor number $N$, timestep length $T_{\text{ts}}$, epoch number $K$, batch size $M\leq NT_{\text{ts}}$.\\
\textbf{Output}: Optimal policy parameter $\boldsymbol{\theta}^{*}_\text{p}$.

\begin{algorithmic}[1] 
\STATE Initialize policy and value network parameter $\boldsymbol{\theta}_\text{p}, \boldsymbol{w}$;
\FOR {iteration $=1,2,\cdots,I$}
\FOR {actor $=1,2,\cdots,N$}
\WHILE {timesteps $<T_{\text{ts}}$}
\STATE Get random $\boldsymbol{x}, \hat{\boldsymbol{x}}$ from $\boldsymbol{X}, \hat{\boldsymbol{X}}$;
\STATE $\boldsymbol{s} \gets (\hat{\boldsymbol{x}} - \boldsymbol{x})\odot\boldsymbol{M}_\text{spatial}$;
\WHILE {$\boldsymbol{s}$ is not terminal state}
\STATE Run policy $\pi_{\boldsymbol{\theta}_\text{old}}$, update state and get reward;
\ENDWHILE
\ENDWHILE
\STATE Compute advantage estimate $A$ (Equation (9));
\ENDFOR
\STATE Optimize $\boldsymbol{\theta}_\text{p}, \boldsymbol{w}$ for $K$ epochs with batch size $M$;
\STATE $\boldsymbol{\theta}_\text{old} \gets \boldsymbol{\theta}_\text{p}$;
\ENDFOR
\STATE \textbf{return} $\boldsymbol{\theta}_\text{p}$.
\end{algorithmic}
\end{algorithm}

\paragraph{Network Architecture.}
The input of the policy network at decision step $t$ is the current state $\boldsymbol{s}_t$.
The network consists of 5 convolutional layers and 3 fully connected (FC) layers.
The kernel size and stride in the time dimension is set to $1$ which makes the network only perform spatial convolutions on each frame.
We use average pooling to sample the overall feature information and to reduce the parameter dimension.
The FC layers compare time domain information of a frame with its surrounding frames 
and it ends with the softmax function which predicts a probability distribution $\boldsymbol{p}_t$ for each frame.

\paragraph{Action Selection.}
In the training process, the action $\boldsymbol{a}_t$ is sampled from a multinomial distribution as
\begin{equation}
a_t = {\rm{Multinomial}}(\boldsymbol{p}_t).
\end{equation}
When making decisions, the agent selects the frame with max probability according to
\begin{equation}
a_t = \mathop{\arg\max}_{a} (\boldsymbol{p}_t). 
\end{equation}

\subsubsection{Policy Optimization}

We train the policy network $\pi_{\boldsymbol{\theta}_\text{p}}(a_t|\boldsymbol{s}_t)$ to optimize the expected sum of rewards
$L(\boldsymbol{\theta}_\text{p})=\mathbb{E}_t[R(\boldsymbol{s}_{1:t})|\boldsymbol{s}_0,\boldsymbol{\theta}_\text{p}]={\sum_{\boldsymbol{s}_{i}}\sum_{a_{i}}\pi_{\boldsymbol{\theta}_\text{p}}(a_i|\boldsymbol{s}_i)r(\boldsymbol{s}_i,a_i)}$.
Here, the expectation $\mathbb{E}$ indicates the empirical average over a finite batch of samples.

We apply PPO algorithm with clipped surrogate objective~\cite{schulman2017proximal}, 
which we found to be more stable and sample efficient than REINFORCE~\cite{williams1992simple}.
We use a value network $V(\boldsymbol{s}_t)$ to approximate the expected future rewards for being in state $\boldsymbol{s}_t$.
The value network has a similar architecture as the policy network but the top FC layer only outputs a single value. 
Denote $\boldsymbol{\theta}_\text{old}$ as the policy parameters before the update, 
and $v_t(\boldsymbol{\theta}_\text{p})$ as the probability ratio $\frac{\pi_{\boldsymbol{\theta}_\text{p}}(a_t|\boldsymbol{s}_t)}{\pi_{\boldsymbol{\theta}_{\text{old}}}(a_t|\boldsymbol{s}_t)}$.
The surrogate objective function that PPO maximizes can be written as
\begin{equation}
\resizebox{.91\linewidth}{!}{$
    \displaystyle
L(\boldsymbol{\theta}_\text{p})=\mathbb{E}_t[{\rm min}(v_t(\boldsymbol{\theta}_\text{p})A_t,{\rm clip}(v_t(\boldsymbol{\theta}_\text{p}),1-\epsilon_\text{p},1+\epsilon_\text{p})A_t)].
$}
\end{equation}
The first term inside $\rm min$, $v_t(\boldsymbol{\theta}_\text{p})A_t$, is the the objective used in Conservative Policy Iteration (CPI) algorithm~\cite{kakade2002approximately} which finds an approximately optimal policy.
The second term ${\rm clip}(v_t(\boldsymbol{\theta}_\text{p}),1-\epsilon_\text{p},1+\epsilon_\text{p})A_t$ modifies the objective by clipping $v_t(\boldsymbol{\theta}_\text{p})$ into the interval $[1-\epsilon_\text{p},1+\epsilon_\text{p}]$.
We take the minimum of the unclipped and clipped objective.
This scheme penalizes changes to the policy that move $v_t(\boldsymbol{\theta}_\text{p})$ away from $1$.
We use the advantage estimator
\begin{equation}
A_t=\delta_t+(\gamma\lambda)\delta_{t+1}+\cdots+(\gamma\lambda)^{(G-t+1)}\delta_{G-1},
\end{equation}
where $G$ is the total number of timesteps when encountering the next terminal state and $\delta_t=r_t+\gamma V(s_{t+1})-V(s_t)$.
In our experiments, we set the hyperparameters $\epsilon_\text{p}=0.2$, $\gamma=0.99$ and $\lambda=0.95$.

During optimization, in each iteration, each of $N$ parallel actors $\pi_{\boldsymbol{\theta}_{\text{old}}}$ collect $T_{\text{ts}}$ timesteps of data 
and compute advantage estimates $A_1,A_2,\cdots,A_{T_{\text{ts}}}$.
Then we construct the surrogate function on these $NT_{\text{ts}}$ timesteps of data, and optimize it with Adam for $K$ epochs.
Algorithm $1$ summarizes the whole procedure of policy optimization.

\subsection{Video Attacking}
There are several algorithms that have been proposed for the black-box attack setting for generating adversarial images.
However, most of them require huge amount of queries for attacking one sample.
Cheng \emph{et al.}~\shortcite{cheng2018query} suggest that this problem can be formulated as an optimization problem, where the objective function can be evaluated by binary search with additional model queries.
Thereby a zeroth order optimization algorithm can be applied to solve the formulation.
They~\shortcite{cheng2019sign} later improved it by directly estimate the sign of gradient at any direction instead of the gradient itself, which enjoys the benefit of single query.
Our attack algorithm is built based on their Sign-OPT algorithm.
\begin{algorithm}[tb]
\caption{Reinforcement learning based sparse targeted video attack}
\label{alg:algorithm2}
\textbf{Input}: Video recognition model $F$, clean video $\boldsymbol{x}$, target class label $y_\text{adv}$, 
target class video set $\hat{\boldsymbol{X}}$, 
pretrained policy parameter $\boldsymbol{\theta}_\text{p}$, 
pretrained value network parameter $\boldsymbol{w}$.\\
\textbf{Parameter}: Sampled examples number $n$, salient region ration $\varphi$, 
attack iteration number $T_\text{ai}$, gradient sampling number $Q$, smoothing parameter $\epsilon_\text{d}$ and step size $\eta$.\\
\textbf{Output}: Adversarial example $x_\text{adv}$.

\begin{algorithmic}[1] 
\STATE Fine-tune the policy network $\pi_{\boldsymbol{\theta}_\text{p}}$ using Algorithm 1 with $\boldsymbol{x}$ and random samples from $\hat{\boldsymbol{X}}$;
\FOR {$i=1,2,\cdots,n$}
\STATE Get random $\hat{\boldsymbol{x}}$ from $\hat{\boldsymbol{X}}$;
\STATE $\boldsymbol{M}_\text{spatial} \gets S(\boldsymbol{x},\varphi) \cup S(\hat{\boldsymbol{x}},\varphi)$;
\STATE Run policy $\pi_{\boldsymbol{\theta}_\text{p}}$ to get temporal sparse mask $\boldsymbol{M}$;
\STATE $\hat{y} \gets F(\boldsymbol{x} + (\hat{\boldsymbol{x}} - \boldsymbol{x})\odot\boldsymbol{M})$;
\IF {$\hat{y} = y_\text{adv}$}
\STATE $\boldsymbol{\theta} \gets \frac{(\hat{\boldsymbol{x}} - \boldsymbol{x})\odot\boldsymbol{M}}{\left\|(\hat{\boldsymbol{x}}-\boldsymbol{x})\odot \boldsymbol{M}\right\|}$;
\STATE Compute $g(\boldsymbol{\theta})$ using the binary search algorithm;
\IF {$MAP(g(\boldsymbol{\theta})\odot\boldsymbol{\theta}) < MAP(g(\boldsymbol{\theta_\text{d}})\odot\boldsymbol{\theta_\text{d}})$}
\STATE $\boldsymbol{M}^*, \boldsymbol{\theta_\text{d}} \gets \boldsymbol{M}, \boldsymbol{\theta}$;
\ENDIF
\ENDIF
\ENDFOR
\FOR {$t = 1,2,\cdots,T_\text{ai}$}
\STATE Sample random $\boldsymbol{u}_1, \boldsymbol{u}_2,...,\boldsymbol{u}_Q$ from Gaussian distribution $\mathcal{N}(0,1)$;
\STATE $ \hat{\boldsymbol{g}} \gets \frac{1}{Q_\text{d}}\sum_{q=1}^{Q_\text{d}} {\rm sign}(g(\boldsymbol{\theta}_\text{d} + \epsilon_\text{d}\boldsymbol{u}_q) -g(\boldsymbol{\theta}_\text{d}))\boldsymbol{u}_q$;
\STATE $\boldsymbol{\theta}_\text{d}\gets \boldsymbol{\theta}_\text{d} -\eta\hat{\boldsymbol{g}}$;
\STATE Evaluate $g(\boldsymbol{\theta}_\text{d})$ using the binary search algorithm;
\ENDFOR
\STATE $\boldsymbol{x}_\text{adv} = \boldsymbol{x} + g(\boldsymbol{\theta}_\text{d}) \cdot \boldsymbol{\theta}_\text{d}$;
\STATE \textbf{return} $\boldsymbol{x}_\text{adv}$.
\end{algorithmic}
\end{algorithm}

\begin{table*}[t]\small
    \centering
    \begin{tabular}{c|c|l|r r r r|r r r r}
    \specialrule{1pt}{0pt}{0pt}
    \multirow{2}{*}{Dataset} & \multirow{2}{*}{Threat Model} & \multirow{2}{*}{Attack Methods} & \multicolumn{4}{c|}{Untargeted Attacks}  & \multicolumn{4}{c}{Targeted Attacks} \cr \cline{4-11} & & &$Q$ & $MAP$& $S$ (\%)& $t$ (min)& $Q$ & $MAP$& $S$ (\%)& $t$ (min) \cr
    \specialrule{1pt}{0pt}{0pt}
    \multirow{8}{*}{UCF-101} &  \multirow{4}{*}{C3D} & OPT & 54313.7 & 3.8807 & 0.00 & 54.01 & 124768.0 & 12.6996 & 0.00 & 64.14\cr
    & &  Heuristic & 49648.8 & 3.8358 & 19.00 & 52.70 & 198685.7 & 10.0651 & 29.50 & 104.31\cr
    & &  Sign-OPT & 13305.7 & 2.9076 & 0.00 & 14.08 & 31639.8 & 8.9471 & 0.00 & 24.71\cr
    & &  Our method & \textbf{4172.9} & \textbf{2.4333} & \textbf{69.63} & \textbf{5.72} & \textbf{12990.5} & \textbf{8.8138} & \textbf{47.85} & \textbf{8.49}\cr \cline{2-11}
    & \multirow{4}{*}{LRCN} & OPT & 12011.7 & 2.8482 & 0.00 & 12.27 & 214718.7 & 18.2769 & 0.00 & 122.53\cr
    & &  Heuristic & 12585.4 & 2.6900 & 15.05 & 13.24 & 265144.3 & 15.1814 & 24.46 & 144.36\cr
    & &  Sign-OPT & 4945.1 & 2.4269 & 0.00 & 5.03 & 32485.1 & 8.9249 & 0.00 & 24.92\cr
    & &  Our method & \textbf{2201.5} & \textbf{2.2892} & \textbf{54.78} & \textbf{2.90} & \textbf{18472.2} & \textbf{8.8039} & \textbf{45.53} & \textbf{11.65}\cr
    \specialrule{1pt}{0pt}{0pt}
    \multirow{8}{*}{HMDB-51} &  \multirow{4}{*}{C3D} & OPT & 21328.3 & 3.4674 & 0.00 & 24.79 & 43975.0 & 12.1758 & 0.00 & 28.51\cr
    & &  Heuristic & 29071.3 & 3.1053 & 18.13 & 33.55 & 103050.8 & 10.1357 & \textbf{45.13} & 53.81\cr
    & &  Sign-OPT & 4273.7 & 2.8127 & 0.00 & 4.62 & 24370.0 & \textbf{8.7348} & 0.00 & 17.79\cr
    & &  Our method & \textbf{2636.6} & \textbf{2.3883} & \textbf{63.63} & \textbf{3.85} & \textbf{16066.9} & 8.7568 & 41.86 & \textbf{9.98}\cr \cline{2-11}
    & \multirow{4}{*}{LRCN} & OPT & 24857.0 & 2.8578 & 0.00 & 29.54 & 40072.9 & 10.9374 & 0.00 & 22.17\cr
    & &  Heuristic & 29340.4 & \textbf{2.5014} & 20.25 & 31.62 & 88080.9 & 8.9667 & 30.09 & 47.42\cr
    & &  Sign-OPT & 4457.9 & 2.5747 & 0.00 & 4.82 & 24063.9 & 8.7354 & 0.00 & 17.43\cr
    & &  Our method & \textbf{2703.6} & 2.5697 & \textbf{67.69} & \textbf{3.81} & \textbf{15742.2} & \textbf{8.7305} & \textbf{45.42} & \textbf{9.78}\cr
    \specialrule{1pt}{0pt}{0pt}
    \end{tabular}
    \caption{Untargeted and targeted attack average results on C3D and LRCN models. For all attack models, $FR$ is $100\%$. $Q$ and $t$ used for pretraining are not counted. In each task, the best performance is emphasized with the bold number.}
    \label{tab.1}
\end{table*}

For simplicity, we only focus on untargeted attack here, while the same procedure can be applied to targeted attack as well.
In our formulation, $\boldsymbol{\theta}_\text{d}$ represents the search direction and $g(\boldsymbol{\theta}_\text{d})$ is the distance from
the given example $\boldsymbol{x}$ to the nearest adversarial example along the direction $\boldsymbol{\theta}_\text{d}$.
The objective function can be written as
\begin{equation}
\resizebox{.89\linewidth}{!}{$
    \displaystyle
 \min_{\boldsymbol{\theta}_\text{d}} \text{ } g(\boldsymbol{\theta}_\text{d})\text{ } {\rm where}  \text{ }g(\boldsymbol{\theta}_\text{d}) = \mathop{\arg\min}_{\lambda>0} (F(\boldsymbol{x}+\lambda\frac{\boldsymbol{\theta}_\text{d}}{\left\|\boldsymbol{\theta}_\text{d}\right\|})\neq \bar{y}).
$}
\end{equation}
It can be evaluated by a binary search procedure locally.
At each binary search step, we query the model $F(\boldsymbol{x}+\lambda\frac{\boldsymbol{\theta}_\text{d}}{\left\|\boldsymbol{\theta}_\text{d}\right\|})$
and determine whether the distance to decision boundary in the direction $\boldsymbol{\theta}_\text{d}$ is greater or smaller than $\lambda$ based on the prediction.
Usually, its directional derivative can be estimated by finite differences 
$\hat{\nabla} g(\boldsymbol{\theta}_\text{d};\boldsymbol{u})=\frac{g(\boldsymbol{\theta}_\text{d}+\epsilon_\text{d}\boldsymbol{u})-g(\boldsymbol{\theta}_\text{d})}{\epsilon_\text{d}}\boldsymbol{u}$,
where $\boldsymbol{u}$ is a random Gaussian vector and $\epsilon_\text{d}$ is a very small positive smoothing parameter.
However, this approach consumes a lot of queries when computing $g(\boldsymbol{\theta}_\text{d}+\epsilon_\text{d}\boldsymbol{u})-g(\boldsymbol{\theta}_\text{d})$.
To improve query complexity, we do not need very accurate values of directional derivative and an imperfect but informative estimation of directional derivative of $g$ which can be computed by a single query is enough for us.
Thus, we can compute the sign of this value by
\begin{equation}
\resizebox{.89\linewidth}{!}{$
\begin{split}
{\rm sign}(g(\boldsymbol{\theta}_\text{d} + & \epsilon_\text{d}\boldsymbol{u}) -g(\boldsymbol{\theta}_\text{d}))= \\
  & \left\{
\begin{aligned}
& 1 && \text{if }  F(\boldsymbol{x}+g(\boldsymbol{\theta}_\text{d})\frac{\boldsymbol{\theta}_\text{d}+\epsilon_\text{d}\boldsymbol{u}}{\left\|\boldsymbol{\theta}_\text{d}+\epsilon_\text{d}\boldsymbol{u}\right\|})\neq \bar{y})\\
& -1 && \text{Otherwise,}
\end{aligned}
\right.
\end{split}
$}
\end{equation}
and estimate the imperfect gradient by sampling $Q_\text{d}$ random Gaussian vectors
\begin{equation}
 \hat{\nabla} g(\boldsymbol{\theta}_\text{d})\approx \hat{\boldsymbol{g}} = \frac{1}{Q_\text{d}}\sum_{q=1}^{Q_\text{d}} {\rm sign}(g(\boldsymbol{\theta}_\text{d} + \epsilon_\text{d}\boldsymbol{u}_q) -g(\boldsymbol{\theta}_\text{d}))\boldsymbol{u}_q .
\end{equation}

When attacking, we first initialize $\boldsymbol{\theta}_\text{d}$ using
\begin{equation}
\boldsymbol{\theta}_\text{d} =  \frac{(\hat{\boldsymbol{x}}-\boldsymbol{x})\odot \boldsymbol{M}}{\left\|(\hat{\boldsymbol{x}}-\boldsymbol{x})\odot \boldsymbol{M}\right\|}.
\end{equation}
We sample $n$ examples of target class video $\hat{\boldsymbol{x}}$ randomly, and choose the one with smallest $g(\boldsymbol{\theta}_\text{d})$.
This helps us to find a good initialization direction and thus get a smaller distortion in the end.
Then we update $\boldsymbol{\theta}_\text{d}$ with the zeroth order optimization and get the optimal direction $\boldsymbol{\theta}^*_\text{d}$.
Finally, we can get the adversarial example $\boldsymbol{x}_\text{adv}$ using
\begin{equation}
\boldsymbol{x}_\text{adv} = \boldsymbol{x} + g(\boldsymbol{\theta}^*_\text{d}) \cdot \boldsymbol{\theta}^*_\text{d}.
\end{equation}
We set $n=100$ in all of the following experiments.

The whole process of our method for targeted attack is described in Algorithm $2$.
The pretrained parameter $\boldsymbol{\theta}_\text{p}$ and $\boldsymbol{w}$ is optimized by Algorithm $1$.
Function $MAP$ computes the mean absolute perturbation which we will define in Equation $(15)$.
For untargeted attack, we skip the fine-tuning step since the pretrained policy network is good enough.

\section{Experiments}
\begin{figure*}
    \centering
    \includegraphics[width = 15.4cm]{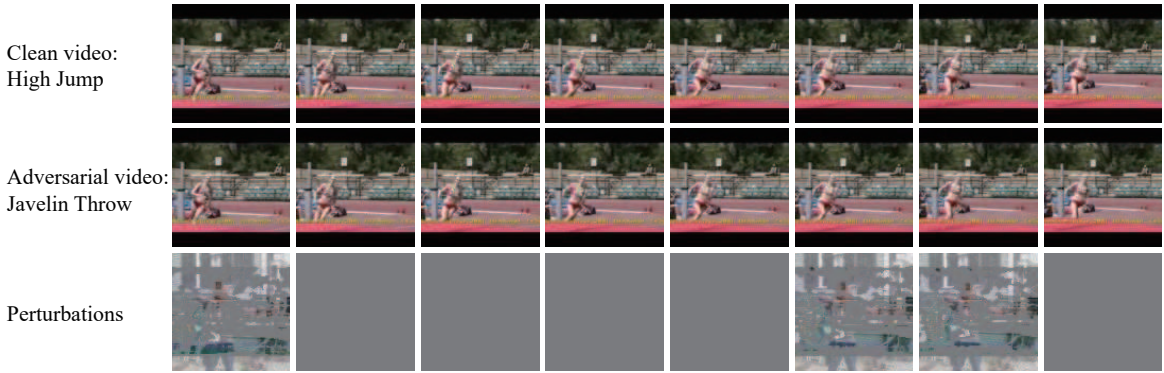}
    \caption{The first 8 frames of an example of adversarial video generated in the untargeted attack by our method. The frames of clean video are shown in the top row and the corresponding adversarial frames are shown in the middle row. The corresponding perturbations are re-scaled into the range of 0-255 and then visualized in the bottom row. The place in grey represents that there is no perturbation.}
    \label{fig:my_label}
\end{figure*}
\subsection{Experimental Setting}
\paragraph{Datasets.}
Similar to~\cite{wei2020heuristic}, we use UCF-101~\cite{soomro2012dataset} and HMDB-51~\cite{kuehne2011hmdb} in our experiments.
UCF-101 contains 13,320 videos collected on YouTube with 101 action categories. 
HMDB-51 contains 7000 clips downloaded from various cinemas distributed in 51 action classes.
We split 70\% of the videos in both datasets as training set and the remaining 30\% as test set. 
We use 16-frame snippets that are uniformly sampled from each video as input samples of threat models and only consider videos that are correctly predicted by the threat model.
During pretraining, we use the training set as both clean video set $\boldsymbol{X}$ and target class video set $\hat{\boldsymbol{X}}$.
When fine-tuning, we choose a video from test set as clean video $\boldsymbol{x}$ and the corresponding training set as target class video set $\hat{\boldsymbol{X}}$.

\paragraph{Threat Models.}
In our experiments, two video recognition models, Convolutional 3D ResNeXt (C3D)~\cite{hara2018can} and Long-term Recurrent Convolutional Networks (LRCN)~\cite{donahue2015long} are used as threat models. 

\paragraph{Metrics.}
We use five metrics to evaluate the performance of our method in various aspects:
1) Fooling ratio ($FR$) is defined as the percentage of adversarial examples that are successfully misclassified.
2) Query number ($Q$) counts the times of queries to finish an attack.
3) Mean absolute perturbation ($MAP$)~\cite{wei2019sparse} denotes the perceptibility score of the adversarial perturbation:
\begin{equation}
MAP = \frac{1}{N_\text{p}}\sum_i|\boldsymbol{\rm r}_i|,
\end{equation}
where $N_\text{p}$ is the number of pixels and $\boldsymbol{\rm r}_i$ is the intensity vector (3-dimensional in the RGB color space).
4) Sparsity ($S$) represents the proportion of pixels with no perturbations versus all the pixels in a specific video, which is defined as 
\begin{equation}
S = 1-\frac{1}{T}\sum_{i}\varphi_i,
\end{equation}
where $\varphi_i$ is the salient region ratio of the corresponding important frame and $T$ is the total number of frames in a video.
5) Time ($t$) counts the time used for finishing an attack.
The computing infrastructures used for running experiments include 8 Nvidia GeForce RTX 2080 Ti GPUs, Intel Xeon E5-2680 v4 CPU, 320 GB memory and Ubuntu 16.04.1.

\subsection{Parameter Setting}
There are several parameters need to be set in our algorithms.
We set most of them by the grid search method. 
Here we only take the parameter salient region ratio $\varphi$ as an example.
Since when $\varphi<0.3$ most initial states will be terminal states, we fix other parameters and set $\varphi$ as $\{0.3, 0.4, 0.5, 0.6, 0.7, 0.8, 0.9, 1.0\}$.
We use 20 randomly sampled videos from the original training set of UCF-101 as validation set to do the experiment.
We find that larger $\varphi$ leads to larger $MAP$ and $Q$, while smaller $\varphi$ causes shorter decision trajectory and less temporal sparsity which also leads to larger $MAP$ and $Q$.
Finally we choose a median $\varphi$ which gets relatively small $MAP$, $Q$ and large $S$ as $\varphi=0.6$.

\subsection{Performance Comparison}
We compare our method with the OPT attack~\cite{cheng2018query} and heuristic attack~\cite{wei2020heuristic}, which we set as the baselines of black-box adversarial attack on video recognition models.
We also compare it with the Sign-OPT attack~\cite{cheng2019sign} as an ablation study for the sparsity in our method.
We are unable to compare our method with~\cite{yan2020sparse} since the code of its model is not publicly available.
The evaluations are performed with two video recognition models on two datasets mentioned above.
For fair comparison, we set the bound of $MAP$ as $MAP=3$ in untargeted attack and $MAP=9$ in targeted attack, according to the median of best performances in the above works.
The iteration of attacking steps will terminate either if $MAP$ reaches the bound or the iteration number exceeds the maximum iteration number $T_\text{ai}=1000$.

Table 1 lists the performance comparisons regarding to both the untargeted attack and the targeted attack on two datasets. The average query numbers $Q$ and time $t$ (min) for pretraining are 17600.5, 11.72 for untargeted attack and 30052.2, 18.68 for targeted attack, respectively. 
The cost of pretraining can easily be offset by no more than 10 times of attack and can be ignored when doing numbers of attacks.

For untargeted attacks, compared to other methods, we have the following observations.
First, our method significantly reduces the query numbers by more than 38\%.
Particularly, it verifies our sparsing strategy's efficiency when comparing to Sign-OPT method with no sparsity.
Second, both our method and Sign-OPT method can reach the $MAP$ bound within a limited iteration number, while OPT method and heuristic method fail on some examples when attacking C3D model.
Third, our method gets better sparsity than heuristic method.
It shows that our agent learns a better frames selection strategy than their method.
Last, by looking into the detailed results, we found that only around 60\% of the queries are spent on attacking, while others are spent on inferring.

For targeted attacks, which usually need more query numbers and perturbations than untargeted attacks due to their difficulty, similar trends are observed as untargeted attacks.
Our method reduces the query numbers by more than 34\% and can always reach the $MAP$ bound within a limited iteration number.
It gets better sparsity than heuristic method on most examples.
We also found that heuristic method does not preform as expectation since many of its generated sparse directions make the attack fall into local optimal solutions.
For summary, our our proposed method achieves the best performance in the majority of test tasks.
It demonstrates the effectiveness and the efficiency of our method.

Figure 1 gives a visualized example of adversarial video generated by our method.
The ground-truth label for the clean video is “High Jump”.
By adding the generated human-imperceptible adversarial perturbations, the model tends to predict a wrong label “Javelin Throw” at the top-1 place.
The generated adversarial perturbations are quite sparse.

\section{Conclusion}
In this work, we proposed an RL based sparse black-box adversarial attack method on video recognition models.
To speed up the attack process and to reduce query numbers, we explored the sparsity of adversarial perturbations.
For temporal domain, we modeled the key frames selection process as an MDP.
The agent in our model continuously explored the differences between videos of distinct classes and received rewards from the threat model.
Combined with policy optimization method, it finally learned an optimized frame selection strategy.
We also used saliency detection to select key regions and Sign-OPT method which only estimates the sign of gradient in zeroth order optimization to further speed up the attack process.
Our algorithm is applicable to multiple threat models and video datasets.
The experimental results demonstrated that our algorithm is more effective and more efficient to generate adversarial examples than previous methods.

The pertinent area of future work is to investigate the black-box attack on video recognition models using fewer queries and time.


\bibliographystyle{named}
\bibliography{ijcai21-id2590}

\end{document}